\documentstyle[12pt,epsf]{article}
\font\elevenbf=cmbx10 scaled\magstep 1
\font\elevenrm=cmr10 scaled\magstep 1
 1

\renewenvironment{thebibliography}[1]
 { \elevenrm
   \begin{list}{\arabic{enumi}.}
    {\usecounter{enumi} \setlength{\parsep}{0pt}
     \setlength{\itemsep}{3pt} \settowidth{\labelwidth}{#1.}
     \sloppy
    }}{\end{list}}
\renewcommand{\thefootnote}{\fnsymbol{footnote}}
\begin{document}
\noindent
\thispagestyle{empty}
\renewcommand{\thefootnote}{\fnsymbol{footnote}}
\begin{flushright}
{\bf TTP94-11\footnote{The complete postscript file of this
preprint, including figures, is available via anonymous ftp at
ttpux2.physik.uni-karlsruhe.de (129.13.102.139) as
/ttp94-11/ttp94-11.ps}}\\
{\bf June 1994}\\
\renewcommand{\thefootnote}{\arabic{footnote}}
\addtocounter{footnote}{-1}
\end{flushright}
\begin{center}
  \begin{large}
RADIATION OF HEAVY QUARKS\\
  \end{large}
  \vspace{0.5cm}
  \begin{large}
   A.H. Hoang${}^{a}$,
   M. Je\.zabek${}^{a,}\,{}^b$, J.H. K\"uhn${}^a$ and T. Teubner${}
^{a,}$\footnote{e--mail: tt@ttpux2.physik.uni-karlsruhe.de}
 \\
  \end{large}
  \vspace{0.1cm}
${}^a$ Institut f\"ur Theoretische Teilchenphysik,Universit\"at Karlsruhe,
    76128 Karlsruhe \\
${}^b$ Institute of Nuclear Physics, Kawiory 26a, PL-30055 Cracow \\
  \vspace{0.7cm}
  {\bf Abstract}\\
\vspace{0.3cm}
\noindent
\begin{minipage}{15.0cm}
\begin{small}
The rate for the production of massive fermions radiated off a pair of
massless fermions is calculated analytically. Combined with the
analytic calculation of the corresponding virtual contribution one
obtains the order $\alpha^2$ correction to the cross section which
is induced by the real and virtual radiation of a pair of massive
fermions  by massless fermions.
Approximations valid for
large and for small masses are derived and shown to agree with
earlier calculations.
\end{small}
\end{minipage}
\end{center}
\vspace{1.2cm}
\noindent 
{\em 1.~Introduction}

A large arsenal of approximations and
expansions has been developed for QCD calculations in
those cases where masses are large or small
compared to the characteristic energy of the problem. There is, however,
only a fairly limited number of three loop problems
with arbitrary nonvanishing masses and energies
where answers can be obtained in
closed analytical form. The results are
first of all useful in their own right; in addition they provide valuable
tests of the approximation methods, strengthening our confidence in
these methods in those cases where no exact result is available.

In this paper the imaginary part of the
three loop amplitude depicted in Fig.\ref{fig1} will be calculated
in closed form.
It corresponds to the rate for the production of a pair of massless
fermions through an external current, with additional radiation of a
(real or virtual) pair of fermions of mass $m$. The calculation will
be performed in the framework of QED; the transition to QCD will be
accomplished at the end of the paper. The calculational technique
is based on the fact that all relevant
integrals can be performed in two steps:
First one evaluates effectively  a two loop diagram;
the result is subsequently integrated with a fairly simple weight
function. This technique has been used previously \cite{KKKS,zdec} to calculate
individually
the contributions of the two fermion and the four fermion cuts
to the imaginary part.
There the limit
$s\gg m^2$ was considered and only logarithmically enhanced or constant terms
were kept.
In \cite{K} the same approach was used to evaluate the vertex corrections
(including of course the appropriate counterterms) for arbitrary $m^2/s$.
In this paper a similar formula is derived for the four fermion cut,
completing thus the ${\cal O}(\alpha^2)$ evaluation of this admittedly rather
simple diagram.

The bulk of the paper is formulated for QED as a reference
theory, with $\alpha$ defined as usual at $q^2=0$ and with fermions of
unit charge. The definition of $m$
is irrelevant in the order under consideration. At the end of the paper
the transition to QCD will be performed and
the $\overline{\rm MS}$ definition of
the coupling constant employed.
\begin{figure}[t]
\begin{center}
\leavevmode
\epsffile[200 360 390 480]{dubbub.ps}
\caption{\label{fig1} {\em
Characteristic Feynman diagram  describing the production
of a pair of massless and a (virtual or real) pair of massive fermions. }}
\end{center}
\end{figure}

\vspace{1cm}\noindent
{\em 2.~Four fermion contribution}

The four fermion cut through the diagram Fig.\ref{fig1} corresponds
essentially to the part of the four fermion production cross section
where the light fermion $q$ with mass zero is coupled to the external
virtual photon or $Z$ boson and the heavy fermion $Q$ with mass $m$
is radiated off the light fermion through a virtual photon.
For the present problem it is convenient to decompose the four fermion
phase space $(q\bar q Q\bar Q)$ integration into the integration over the
three particle phase space $q\bar q (Q\bar Q)$, where $(Q\bar Q)$ denotes
a system of fixed invariant mass $s'$, and a (trivial) two particle phase
space integration for the $(Q\bar Q)$ system. This leaves the final
integration over~$s'$.

The first step corresponds to the calculation of the rate for the decay
of a vector boson of mass $\sqrt s$ into a vector boson of mass
$\sqrt {s'}$ plus a pair of massless fermions $q\bar q$. The two particle
system gives rise to the familiar $Q\bar Q$ threshold factor
$\beta (3-\beta^2)/2$.
One thus arrives at the following integral~\cite{zdec}
\begin{eqnarray}
\frac{\sigma_{q\bar q Q\bar Q}}{\sigma_{q\bar q}({\rm Born})} \, \equiv
\,
R_{q\bar q Q\bar Q} & = & \left(\frac{\alpha}{\pi}\right)^2 \varrho^R\,,
\nonumber \\
\varrho^R & = & \frac{1}{3} \,
\int_{4x}^1 \frac{{\rm d}u}{u} \, (1+\frac{2x}{u}) \,
\sqrt{1-\frac{4x}{u}} \, \times
\nonumber \\
 & & \left\{ \frac{1}{2}\,(1+u)^2\ln^2u + \frac{1}{2}\,(3+4u+3u^2)\ln u +
             \frac{5}{2}\,(1-u^2)
     \right. \nonumber \\
 & &  -\,2(1+u)^2\,[{\rm Li}_2(-u) + \ln(1+u)\ln u + \zeta(2)/2]
     \bigg\}\,,
\end{eqnarray}
where
$u=s'/s$ and $x=m^2/s$.
This integral can be solved in a straightforward way numerically, and after
some effort, also analytically in terms of polylogarithms:
\newpage
\begin{eqnarray}
%
\varrho^R = & &
 \frac{4}{3}\,\left( 1 - 6 x^2 \right) \,
   \left[\frac{1}{2}\,{\rm Li}_3({{1 - w}\over 2}) -
         \frac{1}{2}\,{\rm Li}_3({{1 + w}\over 2})
 \right.  \nonumber \\ & & \left.
\quad+\, {\rm Li}_3({{1 + w}\over {1 + a}}) -
      {\rm Li}_3({{1 - w}\over {1 - a}}) +
      {\rm Li}_3({{1 + w}\over {1 - a}}) -
      {\rm Li}_3({{1 - w}\over {1 + a}})
 \right.  \nonumber \\ & & \left.
\quad +\,\frac{1}{2}\,\ln ({{1 + w}\over {1 - w}})
  \left\{
   \zeta(2) -
   \frac{1}{12}\,\ln^2 (\frac{1+w}{1-w}) +
   \frac{1}{2}\,\ln^2(\frac{a-1}{a+1}) -
   \frac{1}{2}\,\ln (\frac{1+w}{2}) \ln (\frac{1-w}{2})
  \right\}
    \right]
    \nonumber\\ & &
 +\,\frac{1}{9}\,a\,\left( 19 + 46 x \right) \,
  \left(
   {\rm Li}_2({{1 + w}\over {1 + a}}) +
        {\rm Li}_2({{1 - w}\over {1 - a}}) -
        {\rm Li}_2({{1 + w}\over {1 - a}}) -
        {\rm Li}_2({{1 - w}\over {1 + a}}) \right.
    \nonumber\\ & &
  \left.
  \qquad\qquad+\,\ln ({{a - 1}\over {a + 1}})\,\ln ({{1 + w}\over {1 - w}})
    \right)
   \\ & &
 +\,4 \left( {{19}\over {72}} + x + x^2  \right) \,
   \left(
    {\rm Li}_2(-\frac{1+w}{1-w}) - {\rm Li}_2(-\frac{1-w}{1+w})
    -\ln x\, \ln (\frac{1+w}{1-w}) \right)
  \nonumber\\ & &
 +\,7 \left( {{73}\over {189}} + {{74}\over {63}} x +
   x^2\right) \,\ln ({{1 + w}\over {1 - w}})
-\,\frac{1}{3}\left( {{2123}\over {108}} + \frac{2489}{54} x \right) \,w
   \,, \nonumber
\end{eqnarray}
where
\begin{equation}
a=\sqrt{1+4 x}\,, \quad w=\sqrt{1-4 x} \,.
\nonumber
\end{equation}
This formula is the main result of this paper. The function $\varrho^R$ is
evidently zero for $m>\sqrt{s}/2$, corresponding to $x>1/4$. It can be expanded
in the
limit of small masses, $x=m^2/s \ll 1$. Including terms of order $x^3$ one
obtains
\begin{eqnarray}
\varrho^R & = &
 \frac{1}{9}\left[ -\,\frac{1}{2}\ln^3 x - \frac{19}{4}\ln^2 x
   +\left(6\zeta(2)-\frac{73}{3}\right)\ln x+15\zeta(3)
   +19\zeta(2)-\frac{2123}{36}\,\right]
   \nonumber\\ & &
 +\,\frac{4}{3}\,x\left[-\,\frac{3}{2}\ln^2 x-3\ln x+6\zeta(2)-12\,\right]
   \nonumber \\ & &
 +\,\frac{1}{3}x^2\left[\ln^3 x-6\ln^2 x+12\left(-\,\zeta(2)+1\right)\ln x
  -30\zeta(3)+24\zeta(2)\,\right]
  \\ & &
 +\,\frac{2}{9}x^3\left[-\,4\ln^2 x + \frac{28}{3}\ln x
  +12\zeta(2)+\frac{37}{9}\right]  +{\cal O}(x^4)
     \,. \nonumber
\label{rhor}
\end{eqnarray}
The exact result for $\varrho^R$ and approximations including
successively higher orders in $x=m^2/s$ are displayed in Fig.\ref{fig2}a.
The curve including terms up to $m^4/s^2$ already provides an
excellent approximation.

\vspace{1cm}\noindent
{\em 3.~Vertex correction}

The vertex correction has been calculated in \cite{K}. Through the
two particle cut it leads to the following contribution to $R$:
\begin{equation}
\delta R_{q\bar q} =
\left(\frac{\alpha}{\pi}\right)^2 \varrho^V\,,
\end{equation}
with
\begin{eqnarray}
\varrho^V & = &
 \frac{2}{3}\left(1-6x^2\right)\left({\rm Li}_3(A^2) - \zeta(3)-
   2\zeta(2)\ln A+\frac{2}{3}\ln^3 A\right)
   \nonumber\\& &
 +\,\frac{1}{9}\left(19+46x\right)\sqrt{1+4x}\left({\rm Li}_2(A^2)
   -\zeta(2)+\ln^2 A\right)
   \\& &
 +\,\frac{5}{36}\left(\frac{53}{3}+44x\right)\ln x+\frac{3355}{648}
  +\frac{119}{9}x
\,, \nonumber
\end{eqnarray}
where
$A=(\sqrt{1+4x}-1)/\sqrt{4 x}\,.$
The function $\varrho^V$ is simply twice the real part of the
form factor $F_2^{(f)}$ defined in \cite{K} in the region $x=1/4r>0$.
The vertex correction is present
for large and small $m^2/s$ as well. The leading
term in the heavy mass expansion has also been calculated
in \cite{Chet} employing
a completely different technique
\begin{equation}
\varrho^V \approx \frac{s}{m^2}\frac{1}{45}\left(
    \ln\frac{m^2}{s}+\frac{22}{5}\right)\,.
\label{rhovappr}
\end{equation}
As previously discussed in \cite{K} (see Fig.~2 of \cite{K}), the heavy
mass expansion provides an excellent approximation
to the full answer from $m^2 \gg s$ even down to the threshold
$4m^2=s$. This justifies, for example, the
use of eq.(\ref{rhovappr}) for the contribution of virtual top quarks to
the $Z$ decay rate for the full top mass range.
The quality of the expansion is demonstrated in Fig.\ref{figadd}
where the full analytic result for $\varrho^V$ is compared to the
approximation  in the range $s/m^2\leq 4$.
For small $x=m^2/s$, on the other hand, one obtains (terms up to
${\cal O}(x)$can be found also in \cite{K})
\begin{eqnarray}
\varrho^V & = &
\frac{1}{9}\left[\frac{1}{2}\ln^3 x+\frac{19}{4}\ln^2 x+
  \left(-6\,\zeta(2)+\frac{265}{12}\right)\ln x-6\zeta(3)-19\zeta(2)
  +\frac{3355}{72}\right]
  \nonumber\\& &
+\,\frac{4}{3}x\left[\frac{3}{2}\ln^2 x+3\ln x-6\zeta(2)+12\right]
  \nonumber\\& &
+\,\frac{1}{3}x^2
  \left[-\,\ln^3 x+6\ln^2 x+\left(12\zeta(2)-\frac{33}{2}\right)\ln x
  +12\zeta(3)-24\zeta(2)+\frac{39}{2}  \right]
  \\& &
+\,\frac{2}{9}x^3\left[2\ln^2 x-\frac{14}{3}\ln x-8\zeta(2)
  -\frac{1}{3}\right] +{\cal O}(x^4)
\,. \nonumber
\label{rhov}
\end{eqnarray}
\begin{figure}
\begin{center}
\leavevmode
\mbox{}\epsffile[100 300 500 540]{rhor.ps} \\
\mbox{}\epsffile[100 300 500 540]{rhov.ps}
\caption{\label{fig2} {\em
a) The function $\varrho^R$ describing the production of four
fermions in the region $0<x=m^2/s<1/4$. Solid line: exact result;
dashed-dotted line: logarithmic and constant terms only; dotted line:
including $m^2/s$ corrections; dashed line: including $m^2/s$ and
$m^4/s^2$ corrections.
b)~Corresponding curves for $\varrho^V$ describing virtual
corrections. }}
\end{center}
\end{figure}
\begin{figure}
\begin{center}
\leavevmode
\mbox{}\epsffile[100 300 500 540]{rhovlx.ps}
\caption{\label{figadd} {\em
The function $\varrho^V$ describing virtual corrections in the
range $s/m^2<4$ (solid curve) and the approximation eq.(\ref{rhovappr})
(dashed curve).
}}
\end{center}
\end{figure}
In Fig.\ref{fig2}b the function $\varrho^V$ is displayed
in the range from $x=1/4$ down to $1/4 \times 10^{-2}$, together with
approximations including successively higher orders.
Again one concludes that the expansion up to $m^4/s^2$
provides an excellent approximation in the full mass range for nearly
massless quarks as well as close to threshold. Upon adding
$\varrho^R$ and $\varrho^V$ and performing the same approximations as
above, one obtains
\begin{eqnarray}
\label{new}
\varrho^V+\varrho^R & = &
\left[ -\,\frac{1}{4}\ln x+\zeta(3)-\frac{11}{8}\right]
+\,x^2\left[-\,\frac{3}{2}\ln x - 6\zeta(3)+\frac{13}{2} \right]
  \nonumber\\ & &
+\,x^3\left[-\,\frac{4}{9}\ln^2 x+\frac{28}{27}\ln x+
  \frac{8}{9}\zeta(2) +\frac{68}{81} \right] \,+\,{\cal O}(x^4)\,.
\end{eqnarray}
The quadratic and cubic logarithms
compensate as is evident from eqs.(\ref{rhor}) and (\ref{rhov}).
A linear logarithm whose
origin will be discussed in a moment remains.
The sum $\varrho^V+\varrho^R$ is shown in Fig.\ref{fig3}a, together
with the approximations discussed before. Terms proportional $m^2/s$
are absent in the approximations,
whence dotted and dashed-dotted curves coincide.
The absence of these terms has been observed and dicussed in~\cite{ChetKu}
in the context of QCD.

\vspace{1cm}\noindent
{\em 4.~QCD}

It has become customary to formulate the QCD result in the
$\overline{\rm MS}$
renormalization scheme which is better adapted to the limit
$m\rightarrow 0$. To perform this transition (still in the abelian theory)
corrections of order $\alpha$ and $\alpha^2$ must be
redistributed.

With $\alpha$ defined as fine structure constant in the usual
way one finds
\begin{equation}
R=1\,+\,\frac{3}{4}\frac{\alpha}{\pi}\,+\,
\left(\frac{\alpha}{\pi}\right)^2 \left[\left(
\varrho^R +  \varrho^V\right) + ... \, \right]\,.
\end{equation}
The dots indicate other contributions of ${\cal O}(\alpha^2)$ (for example
from two photon exchange or other fermion loops) which are irrelevant
for the present discussion. The fine structure constant is related to
the coupling constant in the $\overline{\rm MS}$ scheme at scale $\mu^2$
through
\begin{equation}
\alpha = \alpha_{\overline{\rm\tiny MS}}(\mu^2)\left(
  1\,+\,\frac{\alpha_{\overline{\rm MS}}(\mu^2)}{\pi}
  \frac{1}{3}\ln\frac{m^2}{\mu^2}
  \right)\,+\,{\cal O}(\alpha^3)
\end{equation}
which implies~\cite{zdec}
\begin{equation}
R=1\,+\,\frac{3}{4}\frac{\alpha_{\overline{\rm MS}}(\mu^2)}{\pi}\,+\,
\left(\frac{\alpha_{\overline{\rm MS}}(\mu^2)}{\pi}\right)^2 \left[\left(
\varrho^R +  \varrho^V+\frac{1}{4}\ln\frac{m^2}{\mu^2}\right)
  + ...\, \right]\,.
\end{equation}
The combination
$
\varrho^R +  \varrho^V+\frac{1}{4}\ln m^2/s
$
thus gives the contribution from the (real plus virtual) radiation of a
massive fermion pair to the total rate or cross section in the
$\overline{\rm MS}$ scheme at scale $\mu^2=s$. It is shown
in Fig.\ref{fig3}b together with the three approximations introduced above.
\begin{figure}
\begin{center}
\leavevmode
\mbox{}\epsffile[100 300 500 540]{sum.ps} \\
\mbox{}\epsffile[100 300 500 540]{sumr.ps}
\caption{\label{fig3} {\em
a) The function $\varrho^R + \varrho^V$ as described in the text.
Solid line: exact result;
dashed-dotted line: logarithmic and constant terms only; dotted line:
including $m^2/s$ corrections; dashed line: including $m^2/s$ and
$m^4/s^2$ corrections.
b) The function $\varrho^R + \varrho^V + \ln x/4$ and
its approximations describing the same result in the $\overline{\rm MS}$
scheme.
}}
\end{center}
\end{figure}

The group theoretical transition from QED (U(1)) to QCD (SU(3))
is easily performed by multiplying the first order correction by
\begin{equation}
\frac{1}{3}{\rm Tr}\left(\frac{\lambda^a\lambda^a}{4}\right) = \frac{4}{3}\,,
\end{equation}
and the second order term by
\begin{equation}
\frac{1}{3}{\rm Tr}\left(\frac{\lambda^a\lambda^b}{4}\right)
\left(\frac{\lambda^a\lambda^b}{4}\right) = \frac{2}{3}\,.
\end{equation}
Adopting as
renormalization scale $\mu^2=s$ one arrives at
\begin{equation}
R_{\rm QCD}=1\,+\,\frac{\alpha_{\rm S}(s)}{\pi}\,+\,
\left(\frac{\alpha_{\rm S}(s)}{\pi}\right)^2 \left[
\left(\frac{2}{3}\right)
\left( \varrho^R +  \varrho^V+\frac{1}{4}\ln\frac{m^2}{s}
\right)   \,
  + ... \,\right]\,.
\end{equation}
The second order correction factor in the small $m$ limit
\begin{eqnarray}
\frac{2}{3}\left(\varrho^R\,+\,\varrho^V\,+
   \,\frac{1}{4}\ln\frac{m^2}{s}\right) & = &
\frac{2}{3}\zeta(3)-\frac{11}{12}
+\,x^2\,\left(-\,\ln x-4\zeta(3)+\frac{13}{3}\right)\\ & &
+\,x^3\,\left(-\,\frac{8}{27}\ln^2 x+\frac{56}{81}\ln x+
    \frac{16}{27}\zeta(2)+\frac{136}{243}\right)\, +\,
  {\cal O}(x^4)\,, \nonumber
\end{eqnarray}
is well behaved. The constant term coincides with the corresponding
result given in~\cite{CKT} (see also~\cite{zdec}),
the quartic mass term with the results of~\cite{CK}.

To summarize: The rate for the production of massive fermions
radiated off a
pair of massless fermions has been calculated analytically. Combined
with the analytic calculation of the corresponding virtual
contribution one obtains the correction to the total cross section of
order $\alpha^2$ which is induced by real and virtual radiation of a
pair of massive fermions. The transition to QCD and to the
$\overline{\rm MS}$ scheme allows the comparison with  approximate
formulae valid for large or small $m^2/s$. Earlier results for
$\alpha_s^2\,m^4/s^2$ corrections have been confirmed. These leading
terms provide an excellent approximation to the full anwer.

\vspace{1cm}\noindent
{\em Acknowledgement:}
We would like to thank K. Chetyrkin for helpful discussions.

\noindent
{\em Note added:} After completion of this calculation we received a
paper by D.E.~Soper and L.R.~Surguladze~\cite{Sur}, where the combination
$ 2/3(\varrho^R+\varrho^V+\frac{1}{4}\ln\frac{m^2}{s})$ has been calculated
numerically.

\vskip 1.5cm
\sloppy
\raggedright
\def\app#1#2#3{{\it Act. Phys. Pol. }{\bf B #1} (#2) #3}
\def\apa#1#2#3{{\it Act. Phys. Austr.}{\bf #1} (#2) #3}
\def\lhc{Proc. LHC Workshop, CERN 90-10}
\def\npb#1#2#3{{\it Nucl. Phys. }{\bf B #1} (#2) #3}
\def\plb#1#2#3{{\it Phys. Lett. }{\bf B #1} (#2) #3}
\def\prd#1#2#3{{\it Phys. Rev. }{\bf D #1} (#2) #3}
\def\pR#1#2#3{{\it Phys. Rev. }{\bf #1} (#2) #3}
\def\prl#1#2#3{{\it Phys. Rev. Lett. }{\bf #1} (#2) #3}
\def\prc#1#2#3{{\it Phys. Reports }{\bf #1} (#2) #3}
\def\cpc#1#2#3{{\it Comp. Phys. Commun. }{\bf #1} (#2) #3}
\def\nim#1#2#3{{\it Nucl. Inst. Meth. }{\bf #1} (#2) #3}
\def\pr#1#2#3{{\it Phys. Reports }{\bf #1} (#2) #3}
\def\sovnp#1#2#3{{\it Sov. J. Nucl. Phys. }{\bf #1} (#2) #3}
\def\jl#1#2#3{{\it JETP Lett. }{\bf #1} (#2) #3}
\def\jet#1#2#3{{\it JETP Lett. }{\bf #1} (#2) #3}
\def\zpc#1#2#3{{\it Z. Phys. }{\bf C #1} (#2) #3}
\def\ptp#1#2#3{{\it Prog.~Theor.~Phys.~}{\bf #1} (#2) #3}
\def\nca#1#2#3{{\it Nouvo~Cim.~}{\bf #1A} (#2) #3}
{\elevenbf\noindent  References \hfil}
\vglue 0.4cm

\end{document}